\documentclass[twocolumn,superscriptaddress,pra,showpacs,aps]{revtex4}

\usepackage{amsmath}
\usepackage{amssymb}
\usepackage{latexsym}
\usepackage{graphicx}
\usepackage[dvips]{color}

\newcommand{\tr}{\textrm{Tr}}

\begin{document}
\title{Initial state dependence of the quench dynamics in integrable quantum systems. \\III.
Chaotic states}
\author{Kai He}
\affiliation{Department of Physics, The Pennsylvania State University, University Park,
Pennsylvania 16802, USA}
\affiliation{Department of Physics, Georgetown University, Washington, DC 20057, USA}
\author{Marcos Rigol}
\affiliation{Department of Physics, The Pennsylvania State University, University Park,
Pennsylvania 16802, USA}

\begin{abstract}
We study sudden quantum quenches in which the initial states are selected to be either eigenstates of
an integrable Hamiltonian that is nonmappable to a noninteracting one or a nonintegrable Hamiltonian,
while the Hamiltonian after the quench is always integrable and mappable to a noninteracting one. By
studying weighted energy densities and entropies, we show that quenches starting from nonintegrable
(chaotic) eigenstates lead to an ``ergodic'' sampling of the eigenstates of the final Hamiltonian,
while those starting from the integrable eigenstates do not (or at least it is not apparent for the
system sizes accessible to us). This goes in parallel with the fact that the distribution of
conserved quantities in the initial states is thermal in the nonintegrable cases and nonthermal in
the integrable ones, and means that, in general, thermalization occurs in integrable systems when the
quench starts form an eigenstate of a nonintegrable Hamiltonian (away from the edges of the
spectrum), while it fails (or requires larger system sizes than those studied here to become
apparent) for quenches starting at integrable points. We test those conclusions by studying the
momentum distribution function of hard-core bosons after a quench.
\end{abstract}
\pacs{02.30.Ik,05.30.-d,03.75.Kk,05.30.Jp}
\maketitle

\section{Introduction}\label{sec:introduction}

The experimental realization of ultracold quantum gases in (quasi-)one-dimensional (1D) geometries
has stimulated much theoretical research on 1D systems \cite{cazalilla_citro_review_11}. In
equilibrium, they are particularly interesting because of the enhanced role played by quantum
fluctuations. In addition, when taken out of equilibrium, they may exhibit unique behavior such as
lack of thermalization \cite{KinoshitaNature06,gring_kuhnert_12}. We note that ultracold gases are
well isolated by the ultra high vacuum in which they are confined, i.e., their dynamics is to a very
good approximation unitary \cite{trotzky_chen_12}. Hence, by lack of thermalization we mean the fact
that, after perturbing the system and letting it evolve, experimental observables relax to
time-independent values that are not described by conventional statistical ensembles. This effect has
been attributed to the proximity to integrable points
\cite{rigol_dunjko_08,RigolPRL09,RigolPRA09,rigol_dunjko_07}, and has motivated an extensive
exploration of the dynamics of integrable systems
\cite{rigol_dunjko_07,rigol_muramatsu_06,cassidy_clark_11,rigol_fitzpatrick_11,he_rigol_12,
cazalilla_06,iucci_cazalilla_09,iucci_cazalilla_10,chung_iucci_12,calabrese_cardy_07,
kollar_eckstein_08,cramer_dawson_08,barthel_schollwock_08,rossini_silva_09,rossini_susuki_10,
mossel_caux_10,fioretto_mussardo_10,calabrese_essler_11,calabrese_essler_12b,calabrese_essler_12a}.

The most common dynamics studied in this context are the ones that occur after so-called sudden
quenches. In a sudden quench, the initial state $|\Psi_I\rangle$ is selected to be an eigenstate
$|\phi_n\rangle$ of an initial Hamiltonian $\hat{H}_I$
($\hat{H}_I|\phi_n\rangle=\epsilon_n|\phi_n\rangle$). Then, at time $\tau=0$, some parameter(s) of
$\hat{H}_I$ is(are) changed instantaneously $\hat{H}_I\rightarrow\hat{H}_F$, and the time evolution
is followed under the final (time-independent) Hamiltonian $\hat{H}_F$
($\hat{H}_F|\psi_\alpha\rangle=E_\alpha|\psi_\alpha\rangle$):
\begin{equation}
    |\Psi(\tau) \rangle = e^{-i\hat H_F \tau/\hbar}| \Psi_I\rangle =
    \sum_{\alpha} C_\alpha e^{-i E_\alpha \tau/\hbar}|\psi_\alpha \rangle,
\end{equation}
where $C_\alpha=\langle\psi_\alpha|\Psi_I\rangle$ are the overlaps of the initial state
with the eigenstates of the final Hamiltonian.

Within this protocol, the time evolution of the expectation value of an observable $\hat{O}$,
$\langle \hat O(\tau) \rangle=\langle\Psi(\tau) |\hat{O} |\Psi(\tau) \rangle$,
can be written as
\begin{equation}\label{eq:timeave}
    \langle \hat O(\tau) \rangle =\sum_\alpha |C_\alpha|^2 O_{\alpha\alpha} +
\sum_{\alpha\neq\beta} {C^*_\alpha}C_\beta e^{i(E_\alpha-E_\beta)\tau/\hbar} O_{\alpha\beta},
\end{equation}
where $O_{\alpha\beta} = \langle \psi_\alpha | \hat O | \psi_\beta \rangle$. If the observable
relaxes to a time-independent value (up to fluctuations that vanish, and revival times that diverge,
in the thermodynamic limit), that value can be computed (up to corrections that vanish in the
thermodynamic limit) by taking the infinite-time average of Eq.~\eqref{eq:timeave}:
\begin{align}\label{eq:diagonal}
   \overline{\langle \hat O(\tau) \rangle} &= \lim_{\tau'\rightarrow \infty} \frac{1}{\tau'}
\int^{\tau'}_0 d\tau\, \langle \hat O(\tau) \rangle \nonumber\\
&=\sum_\alpha |C_\alpha|^2 O_{\alpha\alpha} \equiv \langle \hat O \rangle_{\rm DE},
\end{align}
where we have assumed that there are no degeneracies, or that they can be neglected. The
infinite-time average can be thought of as the prediction of an ensemble, the ``diagonal ensemble''
\cite{rigol_dunjko_07}, where the density matrix is diagonal in the basis of eigenstates of the final
Hamiltonian, i.e., $\hat{\rho}_{\rm DE}=\sum_\alpha |C_\alpha|^2
|\psi_\alpha\rangle\langle\psi_\alpha|$.

Several theoretical studies have shown that in integrable systems, and for observables of interest in
current experiments and models, $\langle \hat O \rangle_{\rm DE}$ is different from the predictions
of conventional statistical mechanics ensembles \cite{rigol_dunjko_07,rigol_muramatsu_06,
cassidy_clark_11,kollar_eckstein_08,calabrese_essler_11,calabrese_essler_12b}. Among the (few-body)
observables that have been studied are the density and momentum distribution function of hard-core
bosons \cite{rigol_dunjko_07,rigol_muramatsu_06,cassidy_clark_11}, the double occupancy of fermions
\cite{kollar_eckstein_08}, and spin-spin correlations in spin chains
\cite{calabrese_essler_11,calabrese_essler_12b}. If one wants to go beyond conventional statistical
mechanics ensembles to take into account that nontrivial conserved quantities $\{\hat{I}_j\}$ (whose
number scales polynomially with system size) are present in integrable systems, and maximizes the
entropy under those constraints \cite{jaynes_57a,jaynes_57b}, then a generalized Gibbs ensemble (GGE)
is obtained \cite{rigol_dunjko_07}. Remarkably, the GGE has been shown to provide the correct
description for the previously mentioned observables (and others) after relaxation following a sudden
quench \cite{rigol_dunjko_07,
rigol_muramatsu_06,cassidy_clark_11,cazalilla_06,iucci_cazalilla_09,iucci_cazalilla_10,chung_iucci_12,
calabrese_cardy_07,kollar_eckstein_08,fioretto_mussardo_10,calabrese_essler_11,calabrese_essler_12b}.
The (grand-canonical) density matrix for the GGE can be written as \cite{rigol_dunjko_07}
\begin{equation}\label{eq:dens GGE}
    \hat \rho_{\rm GGE}=\frac{1}{Z_{\rm GGE}} e^{-\sum_j \lambda_j \hat I_j},
\end{equation}
where $Z_{\rm GGE}=\tr [e^{-\sum_j \lambda_j \hat I_j}]$ is the partition function, $\hat I_j$ are
conserved quantities, and $\lambda_j$ are their corresponding Lagrange multipliers. The latter are
computed using that $\langle\Psi_I|\hat I_j|\Psi_I \rangle = \tr [\hat \rho_{\rm GGE}\, \hat I_j]$.

The lack of thermalization of few-body observables of interest in integrable systems has been
attributed to the failure of the eigenstate thermalization hypothesis
\cite{rigol_dunjko_08,RigolPRL09,RigolPRA09,cassidy_clark_11,DeutschPRA91,SrednickiPRE94}. That is to
say that, within a microcanonical window, the matrix elements $O_{\alpha\beta}$ in an integrable
system fluctuate over a finite range of values. Hence, the outcome of the quench dynamics depends on
which eigenstates of the final Hamiltonian have a finite overlap with the initial state. If the
initial state samples the eigenstates within the microcanonical window in an un-bias way, i.e.,
``ergodically,'' then for that initial state the integrable system will thermalize. This does not
happen for most quenches between integrable systems studied in the literature
\cite{rigol_dunjko_07,rigol_muramatsu_06,
cassidy_clark_11,cazalilla_06,iucci_cazalilla_09,iucci_cazalilla_10,calabrese_cardy_07,kollar_eckstein_08,
fioretto_mussardo_10,calabrese_essler_11,calabrese_essler_12b}. However, it has been shown to occur
for special initial states \cite{rigol_fitzpatrick_11} with particular entanglement properties
\cite{chung_iucci_12}, as well as for initial thermal states at infinite temperature
\cite{he_rigol_12}. In the latter work, finite temperatures were shown to lead to a bias sampling of
the final eigenstates of the Hamiltonian after a quench.

In this work, we study what happens in quenches from initial states that are eigenstates of an
integrable Hamiltonian that is nonmappable to a noninteracting one, while the final Hamiltonian is
integrable and mappable to a noninteracting one. This is of interest as most studies in the
literature have focused on quenches where both the initial and final Hamiltonians are
mappable to noninteracting ones. We also study the case in which the initial state is an eigenstate
of a nonintegrable Hamiltonian and the final Hamiltonian is integrable. As argued in
Ref.~\cite{rigol_srednicki_12}, initial states that are eigenstates of nonintegrable Hamiltonians
(away from the edges of the spectrum \cite{SantosRigolPRE10a,SantosRigolPRE10b}) can lead to
thermalization even though the
final Hamiltonian is integrable. This can be understood as the two-body interactions that break
integrability in models of current interest lead to quantum chaos even in the absence of randomness
\cite{SantosRigolPRE10a,SantosRigolPRE10b,santosPRL2012,santosPRE2012}, similar to what happens in
systems with two-body random interactions \cite{flambaumPRE1996,flambaumPRE1996R,flambaumPRE1997,
flambaumPRE2000}. As a result, the eigenstates of nonintegrable Hamiltonians become random
superpositions of eigenstates of the integrable ones to which the quench is performed, i.e.,
they provide the unbiased sampling required for thermalization after a quench to integrability.
We do not find indications that this happens in quenches between the integrable systems
studied here.

The exposition is organized as follows. In Sec.~\ref{sec:model}, we introduce the model
Hamiltonian and describe the quenches to be studied. The observables
and ensembles of interest are introduced in Sec.~\ref{sec:ensembles}. The remainder sections are
devoted to the discussion of the results for: (i) overlaps between initial states and eigenstates
of the final Hamiltonians (Sec.~\ref{sec:overlap}), (ii) entropies (Sec.~\ref{sec:entropy}),
(iii) conserved quantities (Sec.~\ref{sec:consvquant}), and (iv) momentum distribution functions
(Sec.~\ref{sec:momdist}). A summary of the results is presented in Sec.~\ref{sec:summary}.

\section{Model Hamiltonian and Quenches}\label{sec:model}
We focus our study on 1D lattice hard-core bosons in a box (a system with open boundary conditions)
with nearest-neighbor (NN) hopping $t$ and interaction $V$, and next-nearest-neighbor (NNN) hopping
$t'$ and interaction $V'$. The Hamiltonian reads
{\setlength\arraycolsep{0.0pt}
\begin{eqnarray}\label{eq:hamiltonian}
   && \hat{H} = \sum^{L-1}_{i=1} -t(\hat{b}^\dag_i \hat{b}^{}_{i+1}+\textrm{H.c.}) + V
\left(\hat n_i-\frac{1}{2}\right)\left(\hat n_{i+1}-\frac{1}{2}\right) \nonumber\\
&& + \sum^{L-2}_{i=1} -t'(\hat{b}^\dag_i \hat{b}^{}_{i+2}+\textrm{H.c.}) + V' \left(\hat
n_i-\frac{1}{2}\right)\left(\hat n_{i+2}-\frac{1}{2}\right),\quad\
\end{eqnarray}
}where $\hat{b}^\dag_i(\hat{b}_i)$ is the hard-core boson creation (annihilation) operator,
$\hat{n}_i=\hat{b}^\dag_i\hat{b}_i$ the number operator, and $L$ is the lattice size. In addition to
the usual commutation relations $[\hat{b}_i, \hat{b}^\dag _j]=\delta_{ij}$, $\hat{b}^\dag_i$ and $\hat{b}_i$
satisfy the constraints $\hat{b}^{\dag 2}_i=\hat{b}^2_i=0$, which preclude multiple
occupancy of the lattice sites. In what follows, the NN hopping parameter sets the energy scale ($t=1$),
we take $\hbar=k_B=1$ (where $k_B$ is the Boltzman constant), and the number of particles $N$ is
selected to be $N=L/3$.

For $t'=V'=0$, Hamiltonian (\ref{eq:hamiltonian}) is integrable (it can be mapped onto the spin-1/2
$XXZ$ chain) \cite{cazalilla_citro_review_11}. For $V\neq0$, it is nonmappable to a noninteracting
model. However, if $V=0$,  one can exactly solve Hamiltonian (\ref{eq:hamiltonian}) by first mapping
it onto the spin-1/2 $XY$ chain through the Holstein-Primakoff transformation \cite{Holstein40} and
then onto noninteracting fermions by means of the Jordan-Wigner transformation \cite{Jordan28}, i.e.,
it is mappable to a noninteracting model. Hence, the many-body eigenstates of the fermionic
counterpart are Slater determinants that can obtained from the single-particle Hamiltonian without
the need of diagonalizing the many-body one. Properties of Slater determinants allow one to calculate
off-diagonal correlations of hard-core bosons in polynomial time both in the many-body eigenstates of
the Hamiltonian  \cite{rigol_muramatsu_04_8,rigol_muramatsu_05_13,HePRA11} and in the grand-canonical
ensemble \cite{rigol_05}. The presence of nonzero NNN terms break the integrability of the model and
drive the system to a chaotic regime
\cite{SantosRigolPRE10a,SantosRigolPRE10b,santosPRL2012,santosPRE2012}.

Here, we consider two types of quenches. Quench type I is within integrable models, i.e, we keep
$t'=V'=0$ at all times, and select the initial state to be an eigenstate of Hamiltonian
\eqref{eq:hamiltonian} with $V_I>0$, while the evolution is studied under Hamiltonian
\eqref{eq:hamiltonian} with $V_F=0$. Quench type II is from a nonintegrable model to an integrable
one: Namely, the initial state is selected to be an eigenstate of Hamiltonian \eqref{eq:hamiltonian}
with $t'_I=V'_I\neq0$ while the dynamics is studied under Hamiltonian \eqref{eq:hamiltonian} with
$t'_F=V'_F=0$. In quench type II, we keep $V=0$ at all times. Note that the final Hamiltonian in both
types of quenches is the same. One can fully diagonalize it for large systems by mapping it onto
noninteracting fermions as mentioned before. Since $V\neq0$ and $t'=V'\neq0$ are only present in the
initial Hamiltonian, in what follows we drop the subindex ``$I$'' from $V_I$ and $t'_I=V'_I$.

In order to compare results of quenches from eigenstates of different initial Hamiltonians, we select
the initial states $|\Psi_I\rangle$ such that after the quench the systems have an energy
\begin{equation}\label{eq:energy}
E_I=\langle\Psi_I|\hat{H}_F|\Psi_I\rangle,
\end{equation}
which always corresponds to that of the canonical ensemble (CE) at a fixed temperature $T_{\rm CE}$
(the temperature is taken to be $T_{\rm CE}\approx2$ throughout this work). This means that
$E_I=\text{Tr}\{\hat{\rho}_{\rm CE}\,\hat{H}_F\}$, where $\hat{\rho}_{\rm CE}=e^{-\hat{H}_F/T_{\rm
CE}}/Z_{\rm CE}$ is the CE density matrix, and $Z_{\rm CE}=\text{Tr}\{e^{-\hat{H}_F/T_{\rm CE}}\}$ is
the partition function. The initial states, which are eigenstates of the $t$-$V$ and $t$-$t'$-$V'$
models, i.e., none of which is mappable to a free model, are computed by means of the FILTLAN package
that utilizes a polynomial filtered Lanczos procedure \cite{fang_saad_12}. Since open boundary
conditions are imposed, the only remaining symmetry at 1/3 filling, and for $V\neq2$, is parity.
Here, we restrict our analyses to the even parity sector, the dimension of which is about 1/2 of the
entire Hilbert space. The largest systems we study have $L=24$, whose total Hilbert space dimension
is $D= {24\choose 8}=735\;471$. The even parity subspace for those systems has dimension
$D_{\rm{even}}=367\;983$. This is more than ten times larger than the largest Hamiltonian sector
diagonalized in previous full exact diagonalization studies of similar systems
\cite{RigolPRL09,RigolPRA09,SantosRigolPRE10a,SantosRigolPRE10b,santosPRL2012,santosPRE2012}. In this
work, the combination of utilizing the Lanczos algorithm for computing the initial state, and full
exact diagonalization via the Bose-Fermi mapping for the final Hamiltonian, is what allows us to
perform a complete analysis for larger Hilbert spaces.

\section{Ensembles and Observables}\label{sec:ensembles}

Given the fact that the initial states ($|\Psi_I \rangle$) are pure states, and that the dynamics is
unitary, the von Neumann entropy is zero at all times. In order to characterize our isolated systems,
we make use of the diagonal entropy $S_{\rm DE}$. It is defined as the von Neumann entropy of the
diagonal ensemble \cite{Polkovnikov11},
\begin{equation}\label{eq:entpde}
    S_{\rm DE}=-\sum_\alpha |C_\alpha|^2 \ln\left(|C_\alpha|^2\right),
\end{equation}
and satisfies all thermodynamic properties expected of an entropy \cite{Polkovnikov11}. $S_{\rm DE}$
has been recently shown to be consistent with the entropy of thermal ensembles for generic (nonintegrable)
systems after relaxation \cite{SantosPRL11}.

Similar to the studies in Refs.~\cite{rigol_fitzpatrick_11,he_rigol_12}, we compare the DE with
the entropy predicted by the CE, the grand-canonical ensemble (GE), and the GGE.
Namely, we also compute
\begin{eqnarray}
    S_{\rm CE} &=& \ln Z_{\rm CE}+\frac{E_I}{T_{\rm CE}}, \label{eq:S_CE}\\
    S_{\rm GE} &=& \ln Z_{\rm GE}+\frac{E_I-\mu N}{T_{\rm GE}},\label{eq:S_GE} \\
    S_{\rm GGE} &=& \ln Z_{\rm GGE}+ \sum_j \lambda_j\, \langle\Psi_I|\hat I_j|\Psi_I \rangle \label{eq:S_GGE},
\end{eqnarray}
where $E_I$ was defined in Eq.~\eqref{eq:energy}, and $Z_{\rm CE}$, $T_{\rm CE}$, $N$, and $Z_{\rm GGE}$ were also
defined before. In Eq.~\eqref{eq:S_GE}, the chemical potential and the temperature are computed such that
$E_I=\tr [\hat \rho_{\rm GE}\,\hat H_F]$ and $N=\tr [\hat \rho_{\rm GE}\, \hat N]$, where
$\hat{\rho}_{\rm GE}=e^{-(\hat{H}_F-\mu \hat{N})/T_{\rm GE}}/Z_{\rm GE}$ is the GE
density matrix, $Z_{\rm GE}=\text{Tr}\{e^{-(\hat{H}_F-\mu \hat{N})/T_{\rm GE}}\}$ is
the partition function, and $\hat{N}$ is the total number of particle operator. We note that while the
differences between all those entropies are expected to be finite for any finite system, our goal here is
to understand how those differences behave with increasing system size.

As mentioned before, all our quenches share the same final Hamiltonian ($V=t'=V'=0$), which is
integrable. The conserved quantities $\{\hat{I}_j\}$ for the GGE in that model are taken to be the
occupation number operators of the single particle eigenstates of the fermionic Hamiltonian to which
hard-core bosons can be mapped. From this selection, it follows straightforwardly that the
corresponding Lagrange multipliers can be computed as $\lambda_j=\ln \left[({1-\langle\Psi_I|\hat
I_j|\Psi_I \rangle})/{\langle\Psi_I|\hat I_j|\Psi_I \rangle}\right]$ \cite{rigol_dunjko_07}. Note
that, by construction, the GGE has the same distribution of conserved quantities as the diagonal
ensemble. We also calculate the distribution of conserved quantities predicted by the GE, namely,
$\langle\hat I_j\rangle_\text{GE}=\tr[\hat \rho_{\rm GE}\,\hat I_j]$.

In addition, we study the momentum distribution function of hard-core bosons in the
DE ($\langle\hat m_k\rangle_\text{DE}=\tr[\hat \rho_{\rm DE}\,\hat m_k]$), the
GE ($\langle\hat m_k\rangle_\text{GE}=\tr[\hat \rho_{\rm GE}\,\hat m_k]$), and the
GGE ($\langle\hat m_k\rangle_\text{GGE}=\tr[\hat \rho_{\rm GGE}\,\hat m_k]$). $\hat{m}_k$ is the
diagonal part of the Fourier transform of the one-particle density matrix operator
$\hat{\rho}_{ij}=\hat{b}_i^\dagger\hat{b}^{}_j$,
\begin{equation}
\hat{m}_k=\frac{1}{L}\sum_{i,j=1}^L{e^{\text{i} k(i-j)}}\,\hat{\rho}_{ij}.
\end{equation}
and it was previously studied in systems out of equilibrium with the same $\hat{H}_F$ in
Refs.~\cite{rigol_dunjko_07,rigol_muramatsu_06,cassidy_clark_11}.

\section{overlaps}\label{sec:overlap}

As mentioned before, in integrable systems, the lack of eigenstate thermalization implies that
the distribution of overlaps of the initial state with the eigenstates of final Hamiltonian
$|C_\alpha|^2$ plays a very important role in determining the expectation value of observables
after relaxation [Eq.~\eqref{eq:timeave}]. Therefore, we first study the values of $|C_\alpha|^2$
as a function of $E_\alpha$ in quenches type I and type II and compare them to the weights
of the eigenstates of the final Hamiltonian in the CE, $e^{-E_\alpha /T_{\rm CE}} /Z_{\rm CE}$.

\begin{figure}[!t]
\includegraphics[width=0.475\textwidth]{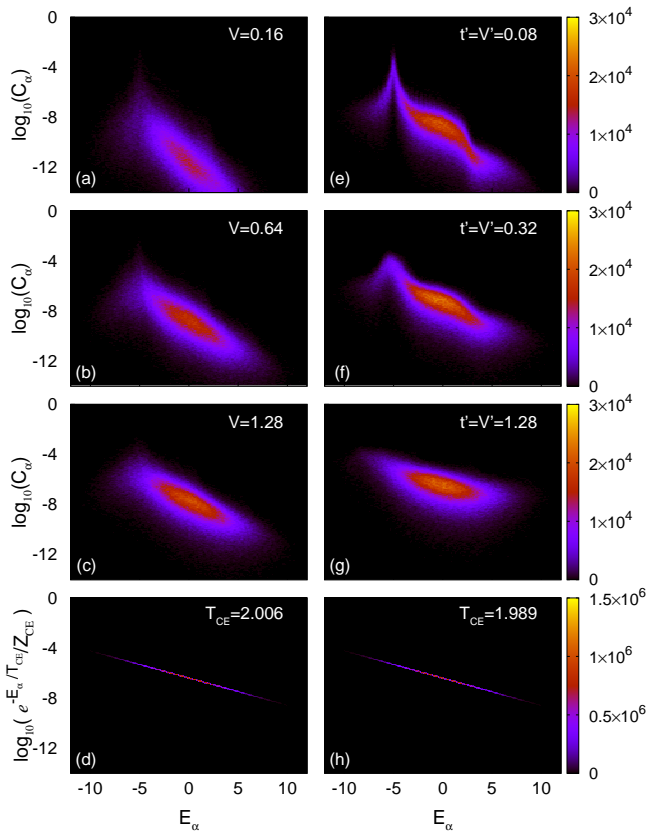}
\caption{(Color online) Coarse-grained plot of the weights in the DE [(a)--(c) and (e)--(g)], and
in the CE [(d) and (h)], $|C_\alpha|^2$ and $e^{-E_\alpha /T_{\rm CE}} /Z_{\rm CE}$,
respectively. (a)--(c) Results for quench type I for different values of $V$. (e)--(g)
Results for quench type II with different values of $t'=V'$. (d) and (h) Weights in the CE,
which correspond to effective temperatures $T_{\rm CE}=2.006$ ($E_I=-4.976$) and
$T_{\rm CE}=1.989$ ($E_I=-5.011$), obtained for the quenches with $V=1.28$ and $t'=V'=1.28$,
respectively. All results are for $L=24, N=8$, and the temperatures are $T_{\rm CE}=2.00\pm0.01$.
The color scale indicates the number of states per unit area in the plot.}\label{fig:coarsegrain}
\end{figure}

In Fig.~\ref{fig:coarsegrain}, we show density plots of the coarse-grained values of $|C_\alpha|^2$
for different type I quenches [Figs.~\ref{fig:coarsegrain}(a)--\ref{fig:coarsegrain}c)] and type II
[Figs.~\ref{fig:coarsegrain}(e)--\ref{fig:coarsegrain}(h)]. The corresponding results for the
canonical weights ($T_{\rm CE}\approx2$) are presented in Figs.~\ref{fig:coarsegrain}(d) and
\ref{fig:coarsegrain}(h). One can see in Fig.~\ref{fig:coarsegrain} that, for all values of $V$ in
quench type I and $t'=V'$ in quench type II, the distribution of weights in the DE is qualitatively
distinct from that in the CE, which shows a simple exponential decay. In the diagonal ensemble, the
weights between eigenstates that are close in energy fluctuate wildly (note the logarithmic scale in
the $y$ axes), and a maximum in the values of $|C_\alpha|^2$ [particularly clear in panels (a), (b),
(e) and (f)] can be seen around the energy of the initial state after the quench $E_I\approx-5.0$.
Hence, in contrast to the canonical ensemble, eigenstates of $\hat{H}_F$ whose energies are close to
$E_I$ usually have the largest weights. This was also seen in Ref.~\cite{rigol_10}.

Another feature that is apparent in Fig.~\ref{fig:coarsegrain} is that while the weights
$|C_\alpha|^2$ decrease rapidly as one moves away from $E_I$, the number of states with those weights
increases dramatically for $E_\alpha>E_I$ due to the increase of the density of states. For this
reason, the appropriate quantity to quantify how different regions of the spectrum contribute to the
observables in the diagonal ensemble (and in the CE) is the weighted energy density function,
\begin{equation}\label{eq:eneden}
    P(E)=\frac{1}{\Delta E}\sum_{|E-E_\alpha|<\Delta E/2} W_\alpha,
\end{equation}
where $W_\alpha$ is $|C_\alpha|^2$ for the DE and $e^{-E_\alpha /T_{\rm CE}} /Z_{\rm CE}$ for the CE.
The summation is limited to the states in a small energy window $|E-E_\alpha|<\Delta E/2$, the width
of which ($\Delta E$) is selected to be the same as for the coarse graining in Fig.\
\ref{fig:coarsegrain}. We have checked that our results are robust for the selected values of $\Delta
E$.

\begin{figure}[!t]
\includegraphics[width=0.475\textwidth]{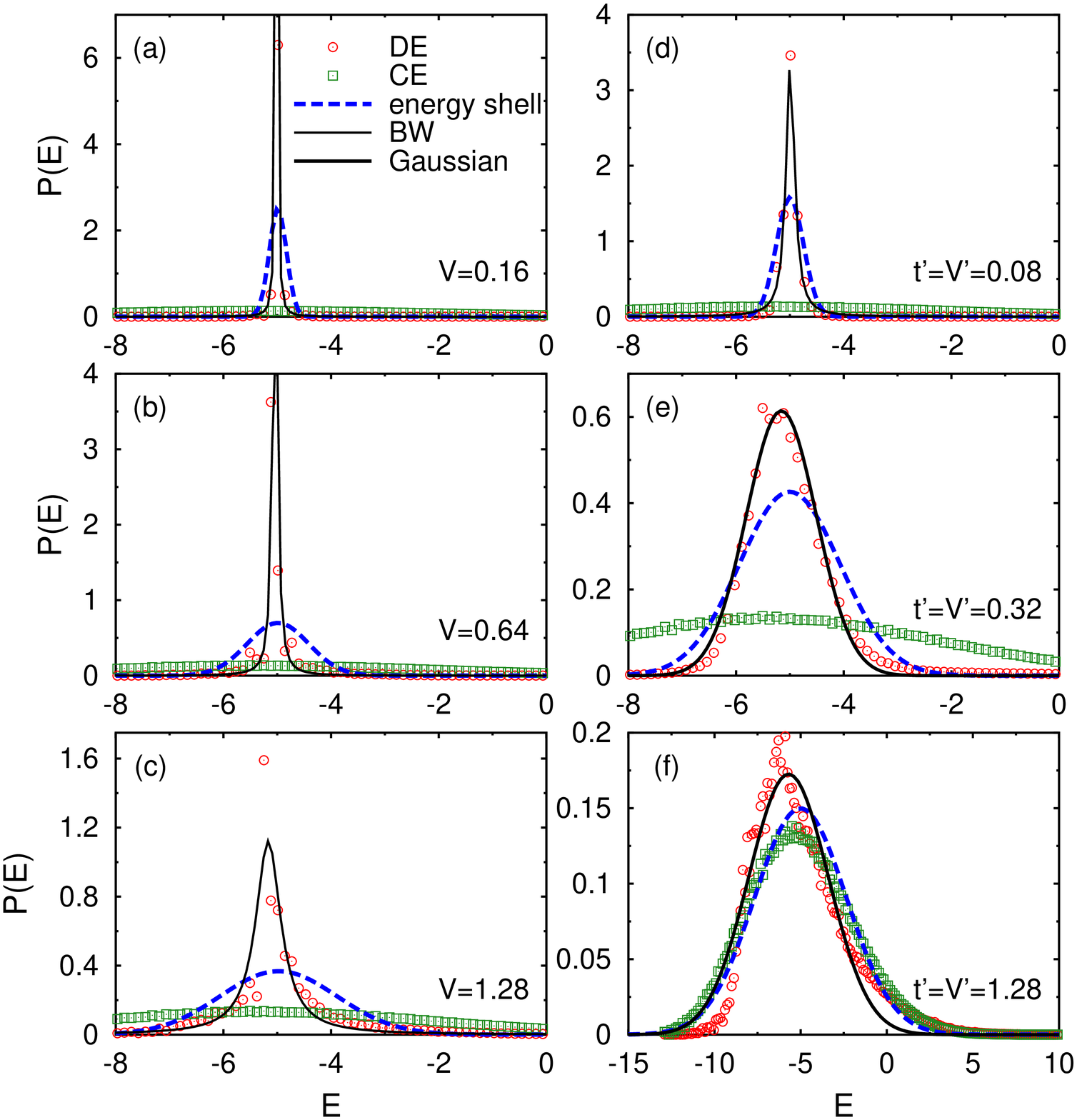}
\caption{(Color online) Weighted energy density functions $P(E)$ for both the DE and the CE. (a)--(c)
Results for quenches type I. (e)--(f) Results for quenches type II. The dashed (blue) lines
depict the energy shell, where (a) $E_I=-4.99,\, \delta E=0.159$, (b) $E_I=-4.99,\, \delta E=0.571$,
(c) $E_I=-4.98,\, \delta E=1.08$, (d) $E_I=-5.00,\, \delta E=0.253$,
(e) $E_I=-5.00,\, \delta E=0.936$, and (f) $E_I=-5.01,\, \delta E=2.66$.
Solid lines show the results of the fits of $P(E)$ to the most appropriate functional form.
Thin lines are fits to a Breit-Wigner function with
(a) $\overline{E}=-5.01,\ \Gamma=0.042$, (b) $\overline{E}=-5.04,\ \Gamma=0.101$,
(c) $\overline{E}=-5.17,\ \Gamma=0.567$, (d) $\overline{E}=-4.99,\ \Gamma=0.187$,
while thick lines are fits to a Gaussian with
(e) $\overline{E}=-5.16,\ \Delta=0.650$, and (f) $\overline{E}=-5.70,\ \Delta=2.31$.
In all cases $L=24$ and $N=8$.
}\label{fig:engdistfunc}
\end{figure}

The weighted energy density functions for the quenches studied in Fig.~\ref{fig:coarsegrain} are
shown in Fig.~\ref{fig:engdistfunc}. There one can see that in all cases, except for the strongest
quenches type II, $P(E)$ in the CE is much broader than that in the DE. For comparison, we have also
plotted the results for the energy shells, which are Gaussians $(\sqrt{2 \pi} \delta
E)^{-1}\exp[-(E-E_I)^2/(2\delta E^2)]$ with the same mean energy $E_I$ and energy width $\delta
E^2=\sum_\alpha |C_\alpha|^2 (E_\alpha-E_I)^2$ as the diagonal ensemble. The energy shell has the
maximal number of eigenstates of the final Hamiltonian that is accessible to the particular initial
state selected. In quenches type I [Figs.~\ref{fig:engdistfunc}(a)-\ref{fig:engdistfunc}(c)], even
for the largest $V$, $P(E)$ in the diagonal ensemble remains narrower than the energy shell. This is
an indication of the lack of ``ergodicity'' of the initial states associated with quench type I. On
the contrary, in quenches type II [Figs.~\ref{fig:engdistfunc}(d)-\ref{fig:engdistfunc}(f)], the
width of $P(E)$ in the diagonal ensemble approaches that of the energy shell with increasing $t'=V'$.
For the strongest quench considered, $t'=V'=1.28$ \cite{note1}, $P(E)$ in the DE and the CE become very close to
each other as well as to the energy shell [Fig.~\ref{fig:engdistfunc}(f)]. The latter shows that
initial states that are eigenstates of nonintegrable Hamiltonians sufficiently distant from an
integrable point fill the energy shell ergodically after a quench to integrability.

To better characterize $P(E)$ as one goes from the weakest to the strongest quenches, we have fitted
it to two different functions: (i) a Breit-Wigner function,
\begin{equation}
P(E) = \frac{1}{2\pi}\frac{\Gamma}{(E-\overline{E})^2+(\Gamma/2)^2},
\end{equation}
and (ii) a Gaussian,
\begin{equation}
P(E) = \frac{1}{\sqrt{2 \pi}\Delta}\exp\left[\frac{-(E-\overline{E})^2}{2\Delta^2}\right],
\end{equation}
where the mean energy $\overline{E}$ and half-width $\Gamma$ of the Breit-Wigner function, as well as
the mean energy $\overline{E}$ and half-width $\Delta$ of the Gaussian, are taken as fitting parameters.

Figures \ref{fig:engdistfunc}(a)-\ref{fig:engdistfunc}(c) show that, as the strength of the
interaction increases in quenches type I, the weighted energy densities in the DE are better
described by Breit-Wigner functions. In quenches type II
[Figs.~\ref{fig:engdistfunc}(d)-\ref{fig:engdistfunc}(f)], on the other hand, weighted energy
densities are described by Breit-Wigner functions for weak quenches and transition to Gaussians as
the strength of the quench increases. The fact that $P(E)$ in the latter quenches approach Gaussians
(and ultimately the energy shell) hints the possibility of observing thermalization in those cases.
This is consistent with the results in Refs.~\cite{santosPRL2012,santosPRE2012}, which considered a
different set of quench protocols and were averaged over different initial states. Note that, in our
results, no average has been introduced.

Using the Breit-Wigner function to fit the weighted energy densities for the quenches type I, as well as
the weakest quench type II, is motivated by analytic results obtained in the random two-body
interaction model \cite{flambaumPRE1996,flambaumPRE1996R,flambaumPRE1997, flambaumPRE2000}, and by
recent numerical results for models similar to those considered here
\cite{santosPRL2012,santosPRE2012}. In Refs.~\cite{santosPRL2012,santosPRE2012}, the weighted energy
densities (which were called strength functions) were computed for the reverse quenches to those
considered here, namely, starting from a state $|\Phi\rangle$ that is an eigenstate
$|\psi_\alpha\rangle$ of an integrable Hamiltonian (in one case, our $\hat{H}_F$) and projecting that
state onto the eigenstates $|\phi_n\rangle$ of the $t$-$V$ and $t$-$t'$-$V'$ Hamiltonians (our
$\hat{H}_I$), i.e., $|\Phi\rangle = \sum_n c_n |\phi_n \rangle$. In our notation, the weighted energy
densities from those works can be written as
\begin{equation}
    P(\epsilon)=\frac{1}{\Delta \epsilon}\sum_{|\epsilon-\epsilon_n|\leq \Delta \epsilon/2} |c_n|^2,
\end{equation}
where the sum runs over states in the spectrum of $\hat H_I$ within a small window of energy
$\Delta\epsilon$. Correspondingly, the energy shell associated with $P(\epsilon)$
is now a Gaussian centered at $\epsilon_S=\sum_n |c_n|^2 \epsilon_n$ whose width is
$\delta \epsilon^2=\sum_n |c_n|^2 (\epsilon_n-\epsilon_S)^2$.

\begin{figure}[!t]
\includegraphics[width=0.475\textwidth]{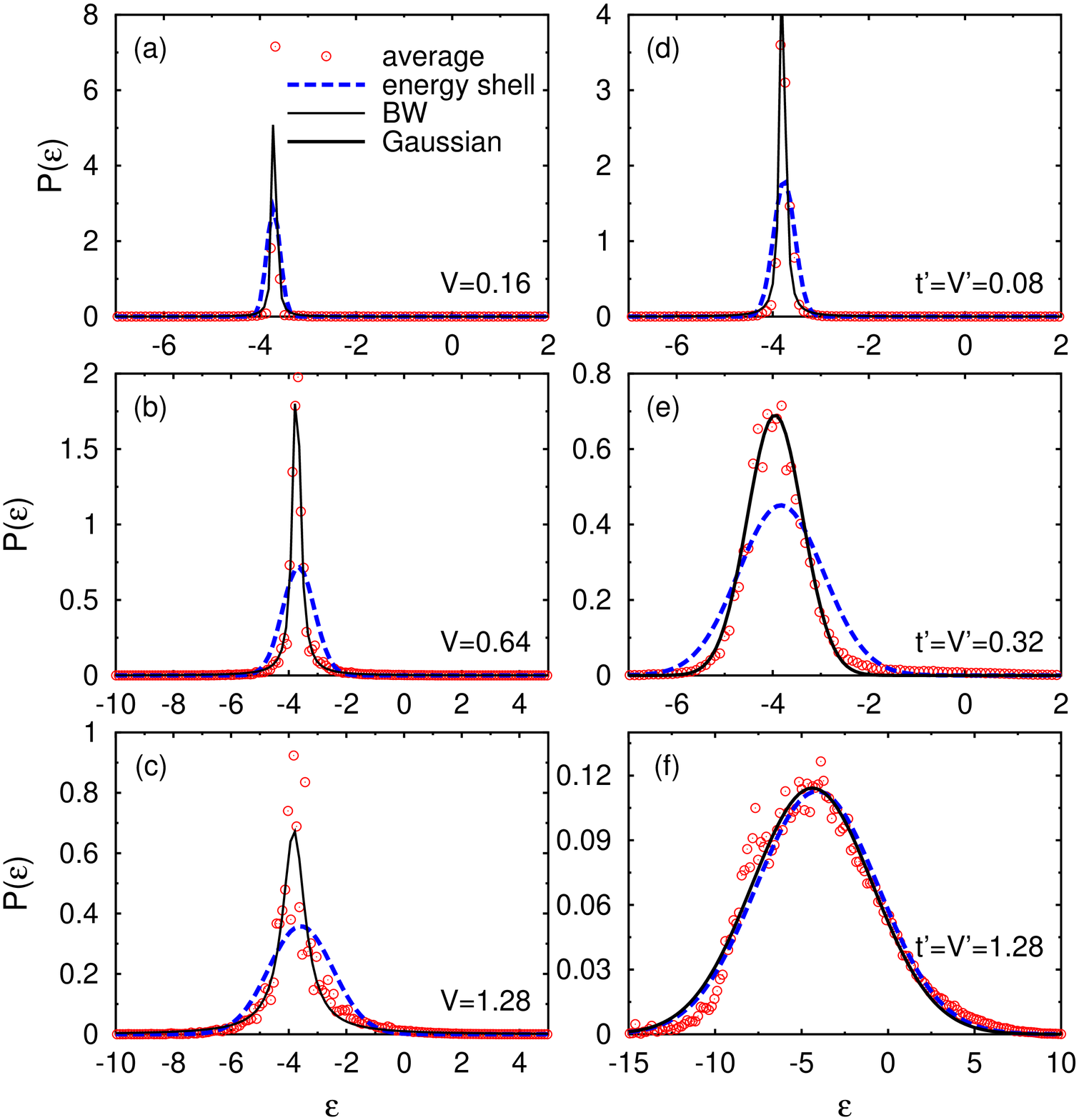}
\caption{(Color online) $P(\epsilon)$ for (a)--(c) quenches type I
and (d)--(f) quenches type II. They were obtained by averaging over the nine even states
$|\psi_\alpha\rangle$ that are closest in energy to $E_I$. Before averaging, we shift $\epsilon_S$
of each distribution to the one obtained for the state $|\psi_\alpha\rangle$ that
is closest in energy to $E_I$. The dashed (blue) lines depict the energy shell, where
(a) $\epsilon_S=-3.72,\, \delta \epsilon=0.139$, (b) $\epsilon_S=-3.66,\, \delta \epsilon=0.557$,
(c) $\epsilon_S=-3.59,\, \delta \epsilon=1.11$, (d) $\epsilon_S=-3.76,\, \delta \epsilon=0.221$,
(e) $\epsilon_S=-3.83,\, \delta \epsilon=0.885$, and (f) $\epsilon_S=-4.10,\, \delta \epsilon=3.54$.
Solid lines show the results of the fits of $P(\epsilon)$ to the most appropriate functional form.
Thin lines are fits to a Breit-Wigner function with
(a) $\overline{\epsilon}=-3.70,\ \Gamma=0.074$, (b) $\overline{\epsilon}=-3.79,\ \Gamma=0.126$,
(c) $\overline{\epsilon}=-3.72,\ \Gamma=0.308$, (d) $\overline{\epsilon}=-3.82,\ \Gamma=0.94$,
while thick lines are fits to a Gaussian with
(e) $\overline{\epsilon}=-3.95,\ \Delta=0.578$, and (f) $\overline{\epsilon}=-4.37,\ \Delta=3.54$.
In all cases $L=18$ and $N=6$.}\label{fig:strfunc}
\end{figure}

For chaotic systems, it was shown in Refs.~\cite{flambaumPRE1996,flambaumPRE1996R,flambaumPRE1997,
flambaumPRE2000,santosPRL2012,santosPRE2012}, that as the interaction strength increases, and the
average strength of the coupling between unperturbed eigenstates becomes of the order of
the average unperturbed level spacing, $P(\epsilon)$ transitions from a Breit-Wigner function to
a Gaussian. This transition also seen numerically in integrable systems \cite{santosPRL2012,santosPRE2012}
through an analysis of states whose energy was in the center of the spectrum, i.e., whose $T_{CE}=\infty$.
In order to make contact with the results in Refs.~\cite{santosPRL2012,santosPRE2012}, we compute
$P(\epsilon)$ for our quenches and in the regions of the spectrum relevant to out work. Note that
since full exact diagonalization is required to obtain all eigenstates $|\phi_n \rangle$, the largest
systems considered here in the calculations of $P(\epsilon)$ have $L=18$ and $N=6$. Finite size
effects are strong for those lattice sizes, so we average the results for $P(\epsilon)$ over
the nine even parity states $|\psi_\alpha\rangle$ that are closest in energy to $E_I$

In Fig.\ \ref{fig:strfunc}, we show results for $P(\epsilon)$ for the same Hamiltonian parameters for
which we previously studied the weighted energy densities. Comparing Fig.\ \ref{fig:strfunc} and
Fig.\ \ref{fig:engdistfunc}, one can see that the results for $P(\epsilon)$ and $P(E)$ are
qualitatively similar. For both types of quenches, $P(\epsilon)$ becomes broader with increasing $V$
(integrable) and $t'=V'$ (nonintegrable). However, when $\hat{H}_I$ is integrable, $P(\epsilon)$
remains better described by Breit-Wigner functions, while a transition from Breit-Wigner to Gaussian
behavior, as well as a filling of the energy shell [better seen in Fig.~\ref{fig:strfunc}(f) than in
Fig.\ \ref{fig:engdistfunc}(f)], is only observed for a nonintegrable $\hat{H}_I$.

Our results for integrable systems are in contrast to those reported in
Refs.~\cite{santosPRL2012,santosPRE2012}, where a transition from Breit-Wigner to Gaussian was also
observed for an interaction quench within integrable systems. This can be attributed to the
combination of two effects. One is the fact that the states $|\psi_\alpha\rangle$ selected here are
far from the middle of the spectrum, as was the case in Ref.~\cite{santosPRL2012,santosPRE2012}.
Within the entire spectrum, the mean level spacing in the middle is the minimal one, i.e., the same
perturbation may couple more states in that region than away from it. If the average level spacing is
the only reason for the difference, then a transition should be seen in our case as one increases the
system size  even further. This is a possibility that cannot be excluded within the present study.

However, there may be another
reason which would make the differences remain in the thermodynamic limit. The result of averaging over
eigenstates in the middle of the spectrum \cite{santosPRL2012,santosPRE2012} leads to an ensemble at
infinite temperature. In that ensemble, the distribution of the quantities that are conserved after the
quench could be featureless. This would result in an un-bias sampling of the eigenstates of the Hamiltonian
after the quench, which may not occur for quenches starting at a finite temperature even if they are
strong quenches. We have found this to be the case in quenches in systems with $V=t'=V'=0$ in which one
starts from a thermal state \cite{he_rigol_12}. Only the initial state at infinite temperature was found
to ergodically sample the eigenstates after the quench, and a finite size scaling analysis showed
that this does not change with increasing system size. We will come back to this point in Sec.~\ref{sec:consvquant},
where the conserved quantities are studied for all quenches considered here.

\section{entropies}\label{sec:entropy}
\begin{figure}[!b]
\includegraphics[width=0.475\textwidth]{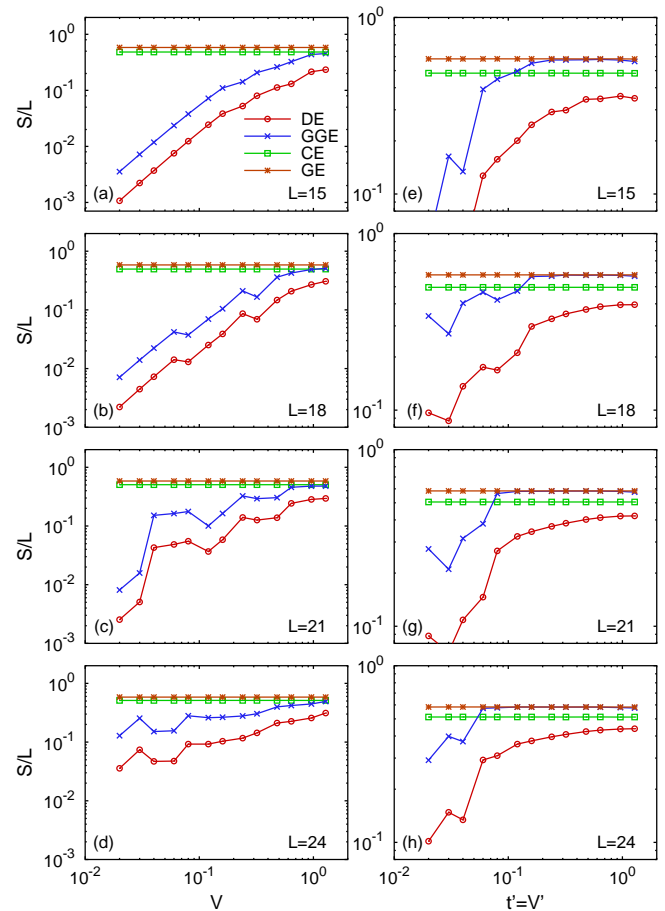}
\caption{(Color online) $S/L$ for the DE, CE, GGE, and GE in systems with $L=15$, 18, 21,
and 24. Results are presented for (a)--(d) quenches type I and (e)--(h) quenches type II.}
\label{fig:entropy}
\end{figure}

The fact that the weighted energy density appears to be Gaussian after a quench to integrability
does not automatically guaranty that thermalization will occur, as it does not mean that
an un-bias sampling has been performed (unless the energy shell is filled). One needs to keep
in mind that a coarse-graining is involved when calculating $P(E)$, where $\Delta E$
[see Eq.~\eqref{eq:eneden}] is much larger than the average level spacing. For example,
in Ref.~\cite{he_rigol_12}, we showed that quenches between integrable systems, which started
from thermal states, led to Gaussian like weighted energy distributions that are not thermal and
do not become thermal with increasing system size. The exception was the initial infinite
temperature state, which does lead to a thermal distribution after a quench (but at
infinite temperature). A qualitatively similar behavior was seen for quenches starting
from special pure states that are ground states of an integrable Hamiltonian \cite{rigol_fitzpatrick_11}.

In order to quantify how the weights evolve in the DE as the system size increases, and how they
compare to the weights in the GGE, CE, and GE, we have computed the entropies of the corresponding
ensembles for all the quenches studied before. In Fig.\ \ref{fig:entropy}, we show all those entropies,
per site, as a function of $V$ for quenches type I [Figs.~\ref{fig:entropy}(a)--\ref{fig:entropy}(d)]
and of $t'=V'$ for quenches type II [Figs.~\ref{fig:entropy}(e)--\ref{fig:entropy}(h)], for four
system  sizes. Since, in all cases, we consider a fixed effective temperature $T_{\rm CE}\approx2$
after the quench, the values of $S/L$ in the CE and the GE are (almost) independent of the quench
parameters.

The first feature that is apparent in Fig.~\ref{fig:entropy} is that, in both types of quenches, the
entropy in the DE and the GGE are very small for weak quenches and increase as system size increases.
The GGE entropy follows (but is always above, as expected from its grand-canonical nature) the DE
entropy. In addition, for each quench, the difference between $S_{\rm DE},\ S_{\rm GGE}$ and $S_{\rm
CE},\ S_{\rm GE}$ is seen to decrease with increasing system size. However, only in quench type II
one can see $S_{\rm GGE}$ to become practically indistinguishable from $S_{\rm GE}$. If one agrees
that the GGE describes observables after relaxation, then the agreement between $S_{\rm GGE}$ and
$S_{\rm GE}$ implies that the observables thermalize in those quenches.

\begin{figure}[!b]
\includegraphics[width=0.475\textwidth]{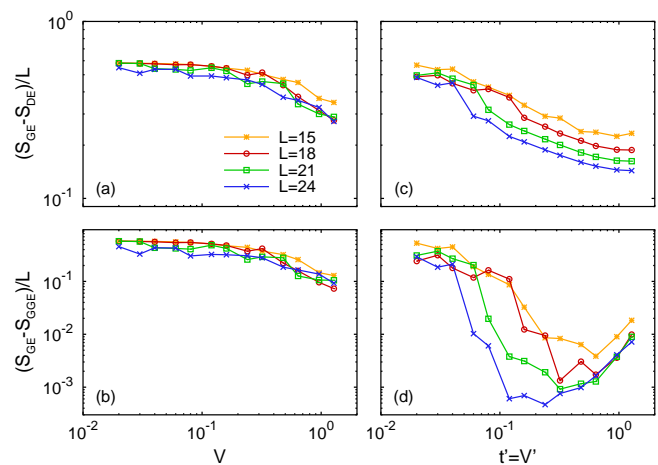}
\caption{(Color online) $(S_{\rm GE}-S_{\rm DE})/L$ as a function of (a) $V$ for quench type I
and (c) $t'=V'$ for quench type II. $(S_{\rm GE}-S_{\rm GGE})/L$ as a function of (b) $V$ for quench
type I and (d) $t'=V'$ for quench type II. Four different system sizes are presented for comparison.
}\label{fig:entropydiff1}
\end{figure}
A better understanding of how the differences between entropies scale with increasing system size can
be gained through Fig.~\ref{fig:entropydiff1}. There we plot $(S_{\rm GE}-S_{\rm DE})/L$ and $(S_{\rm
GE}-S_{\rm GGE})/L$ for different system sizes. Since $(S_{\rm GE}-S_{\rm CE})/L$ vanishes in the
thermodynamic limit, we drop $S_{\rm CE}$ from the remaining analysis. We note that the analysis in
Fig.~\ref{fig:entropydiff1} reflects small fluctuations in $S_{\rm GE}$ (unnoticeable in
Fig.~\ref{fig:entropy}) that occur because $E_I$ is close but not identical in all initial states.

Figure \ref{fig:entropydiff1} shows that, with increasing system size, a decrease of $(S_{\rm
GE}-S_{\rm DE})/L$ and a vanishing of $(S_{\rm GE}-S_{\rm GGE})/L$ is apparent only in quenches type
II, which, once again, indicates that thermalization will occur in those quenches. Furthermore, the
value of $t'=V'$ at which an abrupt reduction of the differences $(S_{\rm GE}-S_{\rm DE})/L$ and
$(S_{\rm GE}-S_{\rm GGE})/L$ occurs decreases as the system size increases, suggesting that in the
thermodynamic limit an infinitesimally small quench type II will lead to thermalization. We should
add that, in Fig.~\ref{fig:entropydiff1}(d), the slight upturn (note the logarithmic scale in the $y$
axes) of $(S_{\rm GE}-S_{\rm GGE})/L$ for the strongest quenches is related to the skewing of the
weighted energy density seen in Fig.~\ref{fig:engdistfunc}. Its effect is imperceptible in
Fig.~\ref{fig:entropydiff1}(c), as $(S_{\rm GE}-S_{\rm DE})/L$ is more than an order of magnitude
larger than $(S_{\rm GE}-S_{\rm GGE})/L$. From the evolution of the results with increasing size, we
expect this upturn to disappear in the thermodynamic limit. For quenches type I, the results in
Figs.~\ref{fig:entropydiff1}(a) and \ref{fig:entropydiff1}(b) show that $(S_{\rm GE}-S_{\rm DE})/L$
and  $(S_{\rm GE}-S_{\rm GGE})/L$, respectively, either remain finite in the thermodynamic limit or
vanish very slowly with increasing system size. If the latter were the case, larger system sizes are
needed to conclusively see a decrease of $(S_{\rm GE}-S_{\rm GGE})/L$.

\section{Conserved Quantities}\label{sec:consvquant}

As mentioned before, the presence of nontrivial sets of conserved quantities make integrable
systems different from nonintegrable ones. Hence, an understanding of whether a particular
initial state leads to thermalization can also be gained from analyzing how the quantities
that are conserved after the quench behave in the initial state. If the distribution of conserved
quantities is identical to the one in thermal equilibrium with energy $E_I$ after the quench,
then the initial state can provide an un-bias sampling of the thermal one and lead to
thermalization. Two particular examples in which that happens, in quenches between
integrable systems, were discussed in Refs.~\cite{rigol_fitzpatrick_11,he_rigol_12}. However,
in those examples, the distributions of conserved quantities were featureless and corresponded
to systems that were at infinite temperature after the quench.

\begin{figure}[!t]
\includegraphics[width=0.475\textwidth]{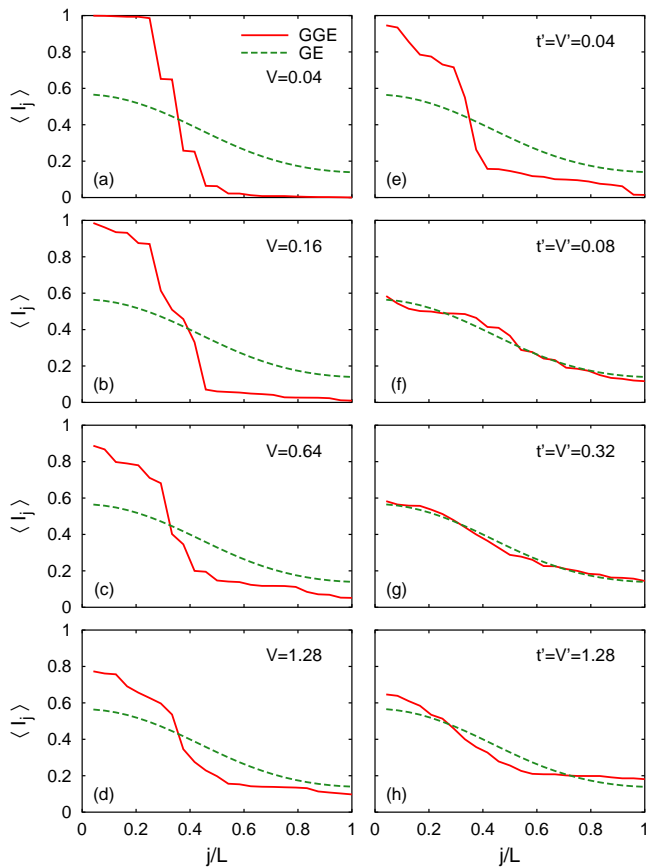}
\caption{(Color online) Distribution of conserved quantities
$\langle\Psi_I|\hat I_j|\Psi_I\rangle$ in the initial state (same as the GGE) and in the GE.
Results are presented for (a)--(d) quenches type I, and (e)--(h) quenches type II.
In all cases, $L=24$ and $N=8$.
}\label{fig:consvquant}
\end{figure}

In Fig.~\ref{fig:consvquant}, we show the distribution of conserved quantities for various initial
states for quenches type I [Figs.~\ref{fig:consvquant}(a)--\ref{fig:consvquant}(d)] and type II
[Figs.~\ref{fig:consvquant}(e)--\ref{fig:consvquant}(h)] in systems with $L=24$, and compare them to
the distribution of conserved quantities in the GE (which is almost identical in all quenches as
$E_I$ is very close in all of them). That figure shows that, in quenches type I, there are large
differences between $\langle\Psi_I|\hat I_j|\Psi_I\rangle$ and the distribution of conserved
quantities in the GE. In contrast, in quenches type II, $\langle\Psi_I|\hat I_j|\Psi_I\rangle$ and
the GE results approach each other and become very similar as $t'=V'$ increases.

In order to have a more quantitative understanding of the difference between the conserved quantities
in the initial state (in the GGE) and in the GE, as the system size increases, we calculate the
relative integrated difference defined as
\begin{equation}\label{eq:intdiff}
\Delta I = \frac{\sum_j |\langle\Psi_I|\hat I_j|\Psi_I\rangle-\langle\hat I_j \rangle_{\rm GE}|}
{\sum_j \langle\Psi_I|\hat I_j|\Psi_I\rangle}.
\end{equation}

\begin{figure}[!t]
\includegraphics[width=0.475\textwidth]{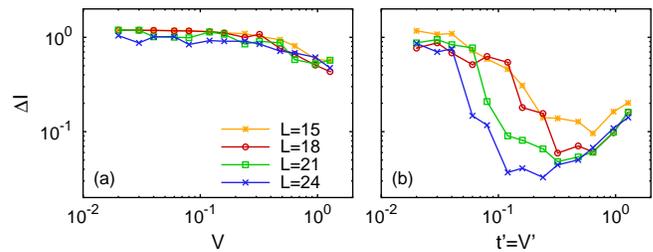}
\caption{(Color online) Relative integrated difference of the conserved quantities in the
initial state (in the GGE) and the GE, $\Delta I$, as a function of (a) $V$ for quench type I and
(b) $t'=V'$ for quench type II. Results are presented for four system sizes.}\label{fig:intdiff}
\end{figure}

Results for $\Delta I$ in both types of quenches, and for different systems sizes, are presented in
Fig.~\ref{fig:intdiff}. They make evident that, as the system size increases in quenches type II, the
conserved quantities in the initial state converge to those predicted in thermal equilibrium [as in
Fig.~\ref{fig:entropydiff1}(d), the upturn seen in Fig.~\ref{fig:intdiff}(b) for large values of
$t'=V'$ is expected to disappear with increasing system size]. No such clear tendency is seen for
quenches type I. We note that the results for $\Delta I$ are qualitatively similar to those obtained
for $(S_{\rm GE}-S_{\rm GGE})/L$ in Fig.~\ref{fig:entropydiff1}. This can be understood as the
entropy in the GGE and in the GE can be written in terms of the occupation of the single-particle
eigenstates in each case:
\begin{equation}
 S=-\sum_{j=1}^L [(1-I_j) \ln(1-I_j)+I_j\ln I_j],
\end{equation}
where $I_j=\langle\Psi_I|\hat I_j|\Psi_I\rangle$ for the GGE and
$I_j=\langle \hat I_j\rangle_{\textrm{GE}}$ for the GE.

The results obtained for quenches type II imply that they will lead to thermalization
in integrable systems. This occurs even though the distribution
of conserved quantities is a nontrivial one (it is not flat, which was the case in
Refs.~\cite{rigol_fitzpatrick_11,he_rigol_12}). Beyond its implication for the quantum
dynamics of integrable systems, one could think of using this information to learn about
complicated many-body systems. It implies, e.g., that if we know the kinetic energy of a
chaotic strongly correlated fermionic system in equilibrium, we could automatically calculate
its momentum distribution by computing the momentum distribution function of a system that is
in thermal equilibrium in the noninteracting limit with the same energy as the kinetic
energy of the strongly correlated one.

\section{momentum distributions}\label{sec:momdist}

In this section, we test whether the conclusion reached in the previous sections, that quenches
type I do not exhibit a clear tendency to thermalize with increasing system size while quenches
type II do, holds for an observable, the momentum distribution function $m_k$.

\begin{figure}[!t]
\includegraphics[width=0.475\textwidth]{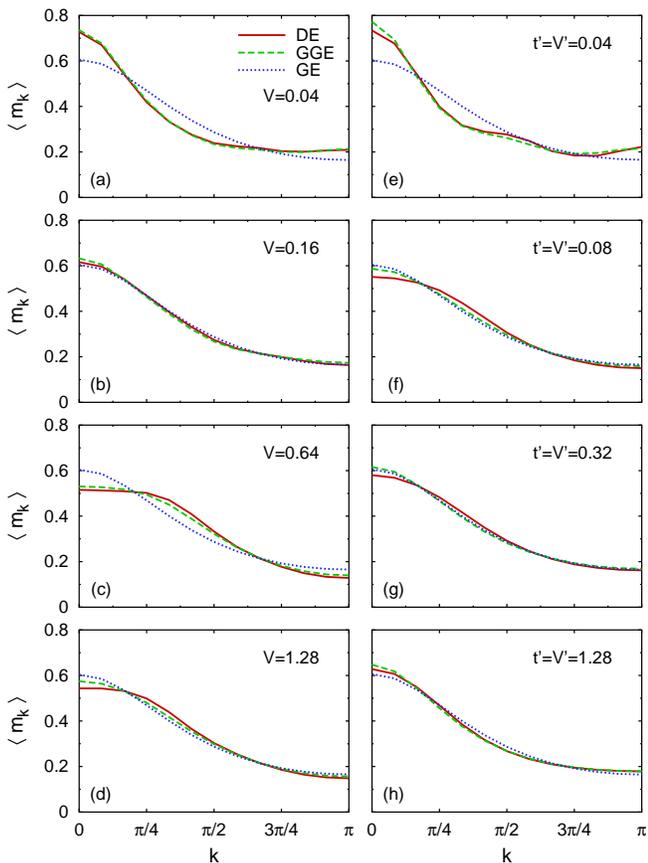}
\caption{(Color online) Momentum distribution functions in the DE, GGE, and GE.
Results are presented for (a)--(d) quenches type I, and (e)--(h) quenches
type II. In all cases, $L=24$ and $N=8$.
}\label{fig:momentum}
\end{figure}

In Fig.~\ref{fig:momentum}, we show the momentum distribution  function obtained within the diagonal
ensemble for various initial states for quenches type I
[Figs.~\ref{fig:momentum}(a)--\ref{fig:momentum}(d)] and type II
[Figs.~\ref{fig:momentum}(e)--\ref{fig:momentum}(h)], and compare them to the momentum distribution
functions predicted by the GGE and the GE \cite{note2}. That figure shows that while the GGE results closely
follow the ones in the DE for all quenches, the GE results are only consistently closer to the DE
ones as one increases $t'=V'$ in quenches type II
[Figs.~\ref{fig:momentum}(f)--\ref{fig:momentum}(h)].

Once again, in order to be more quantitative, we compute the integrated differences between
the GE and DE results, namely
\begin{equation}\label{eq:momentum}
\Delta m_{\rm DE} = \frac{\sum_k |\langle\hat m_k \rangle_{\rm GE}-\langle\hat m_k \rangle_{\rm DE}|}
{\sum_k \langle\hat m_k \rangle_{\rm GE}}.
\end{equation}

\begin{figure}[!b]
\includegraphics[width=0.475\textwidth]{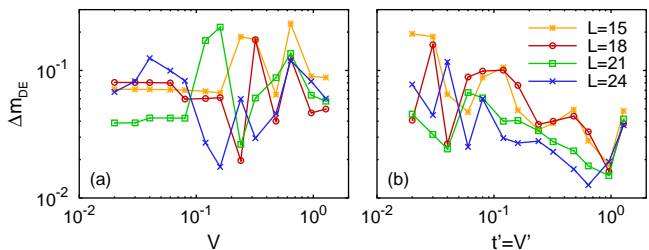}
\caption{(Color online) Relative integrated difference $\Delta m_{\rm DE}$ as a function of
(a) $V$ for quench type I and (b) $t'=V'$ for quench type II.
Results are presented for four system sizes.}\label{fig:momdiff}
\end{figure}
Results for $\Delta m_{\rm DE}$ are presented in Fig.~\ref{fig:momdiff}(a) for quenches type I and in
Fig.~\ref{fig:momdiff}(b) for quenches type II. In the former figure, no consistent trend is seen in
the $\Delta m_{\rm DE}$ with increasing system size. In contrast, in Fig.~\ref{fig:momdiff}(b) one
can see that, for the three largest system sizes, $\Delta m_{\rm DE}$ decreases steadily with
increasing $L$ for $t'=V'$ between 0.1 and 1.  For $t'=V'>1$, but finite, we expect  that
thermalization will also occur, with the upturn seen in Fig.~\ref{fig:momdiff}(b) for $t'=V'>1$
moving to larger values of $t'=V'$ as the system size increases \cite{rigol_srednicki_12}.

The results obtained for the momentum distribution function are
in agreement with what was expected from our previous analysis, and indicates that quenches
type II lead to thermalization, while quenches type I do not lead to thermalization
or require much larger system sizes to observe an approach to the thermal predictions.

\section{summary}\label{sec:summary}

We have studied quenches to an integrable Hamiltonian mappable to free fermions, in which the initial
state is selected to be either an eigenstate of an integrable Hamiltonian that is nonmappable to a
noninteracting one (quench type I) or an eigenstate of a nonintegrable Hamiltonian (quench type II).
By studying weighted energy densities and entropies, we have found no clear evidence that quenches
type I, at least within our Hamiltonians of interest and away from the middle of the spectrum, lead
to an un-bias sampling of the eigenstates of the final Hamiltonian. Quenches type II, on the other
hand, are found to provide such an un-bias sampling. Furthermore, an analysis of different systems
sizes indicates that, in the thermodynamic limit, an infinitesimal quench type II will lead to
thermalization. Here, an important requirement to keep in mind is that the initial state must be away
from the edges of the spectrum. This is because, for systems with two-body interactions in the
absence of randomness, chaotic eigenstates can only be found away from the edges of the spectrum
\cite{SantosRigolPRE10a,SantosRigolPRE10b}.

We have also shown that, in the initial state, an analysis of the distribution of the quantities
that are conserved after the quench provides an understanding of why thermalization occurs in
one type of quenches while it fails in the other one. In quenches type I, that distribution remains
different from, or approaches very slowly with increasing systems size, the one in thermal equilibrium
after the quench. This implies that an un-bias sampling of the eigenstates of the final Hamiltonian
does not occur or takes very large systems to be discerned. The opposite is true for quenches
type II. Hence, quenches type II provide a consistent way of creating initial states that have
the appropriate distribution of conserved quantities so that thermalization can occur after a quench to
integrability. Special initial states for which this occurred in quenches type I were discussed in
Refs.~\cite{rigol_fitzpatrick_11,he_rigol_12}. However, the distribution of conserved quantities
in those cases was (trivially) flat corresponding to infinite temperature systems after the quench.

Finally, by studying the momentum distribution function in quenches type I and type II, we have
shown that the conclusions reached on the basis of the results for the energy distributions, entropies,
and conserved quantities hold. Namely, we have found no indications that quenches type I lead to
the thermalization of this observable while quenches type II do result in thermal behavior.

\begin{acknowledgments}
This work was supported by the U.S. Office of Naval Research. We are grateful to F. M. Izrailev
and L. F. Santos for useful comments on the manuscript.
\end{acknowledgments}


\begin{thebibliography}{50}
\bibitem{note1} We are interested in understanding what happens in the thermodynamic limit for
    quenches that may have arbitrarily large but finite [$O(L^0)$] values of $V$ or $t'=V'$. This
    ensures that the ratio between the width of the weighted energy density distribution after the
    quench and the width of the full energy spectrum vanishes in the thermodynamic limit
    \cite{rigol_dunjko_08}. Hence, here we are not concerned with what happens for a finite system as
    $V$ or $t'=V'$ is increased to arbitrarily large values.
\bibitem{note2} We have also computed the momentum distribution functions within the CE. Those results,
as well as their differences with the ones obtained within the DE and the GGE, are not only qualitatively
but also quantitatively very similar to the ones discussed here for the GE. This is expected from the finite
size scaling analysis of this observable in the CE and the GE presented in Ref.~\cite{rigol_05}.
\end{thebibliography}
\end{document}